%% file: bare_jrnl_new_sample4.tex
\documentclass[lettersize,journal]{IEEEtran}
\usepackage{amsmath,amsfonts}
\usepackage{pifont}
\usepackage{algorithmic}
\usepackage{algorithm}
\usepackage{array}
\usepackage[caption=false,font=normalsize,labelfont=sf,textfont=sf]{subfig}
\usepackage{textcomp}
\usepackage{stfloats}
\usepackage{url}
\usepackage{verbatim}
\usepackage{graphicx}
\usepackage{cite}
\usepackage{multirow}
\usepackage{graphicx}
\usepackage{booktabs}
\usepackage{tabularx}
\usepackage{xspace} 
\usepackage[flushleft]{threeparttable}
\usepackage{chemformula}
\usepackage{soul}
\usepackage{adjustbox}
\usepackage{bm}

\usepackage[hidelinks]{hyperref}
\begin{document}

\title{ApSense: Data-driven Algorithm in PPG-based Sleep Apnea Sensing}

\author{Tanut~Choksatchawathi$^{\dag}$, Guntitat~Sawadwuthikul$^{\dag}$,  Punnawish~Thuwajit, Thitikorn~Kaewlee, Thee Mateepithaktham, Siraphop~Saisaard, Thapanun~Sudhawiyangkul, Busarakum~Chaitusaney$^{*}$, Wanumaidah~Saengmolee$^{*}$ and Theerawit~Wilaiprasitporn$^{*}$, \IEEEmembership{Senior Member, IEEE}
\thanks{\textit{($^{\dag}$T.~Choksatchawathi and G.~Sawadwuthikul equally contributed to this work.) ($^{*}$co-corresponding authors: B.~Chaitusaney, W.~Saengmolee, T.~Wilaiprasitporn).}}
\thanks{T.~Choksatchawathi, G.~Sawadwuthikul, T.~Kaewlee, T.~Sudhawiyangkul, and T.~Wilaiprasitporn are with Vidyasirimedhi Institute of Science \& Technology (VISTEC), Rayong 21210, Thailand and also with Sense AI Company Limited, Rayong 21210, Thailand (e-mail: theerawit.w@vistec.ac.th).}
\thanks{P.~Thuwajit is with the University of Wisconsin Madison, Madison, Wisconsin, U.S..}
\thanks{T.~Mateepithaktham is with the Faculty of Medicine Siriraj Hospital, Mahidol University, Bangkok 10700, Thailand}
\thanks{W.~Saengmolee and S.~Saisaard is with Vidyasirimedhi Institute of Science \& Technology (VISTEC), Rayong 21210, Thailand}
\thanks{B.~Chaitusaney is with Excellence Center for Sleep Disorders, King Chulalongkorn Memorial Hospital, Thai Red Cross Society, Bangkok, Thailand}
\thanks{Copyright (c) 2024 IEEE. Personal use of this material is permitted. However, permission to use this material for any other purposes must be obtained from the IEEE by sending a request to pubs-permissions@ieee.org.}}
\markboth{Journal of \LaTeX\ Class Files,~Vol.~14, No.~8, August~2021}%
{Shell \MakeLowercase{\textit{et al.}}: A Sample Article Using IEEEtran.cls for IEEE Journals}


\maketitle

\begin{abstract}



Detecting obstructive sleep apnea (OSA) is essential for diagnosing and managing sleep health. Traditionally, this involves clinical settings with hardly accessible processes. We propose that automated detection of OSA events is achievable using features extracted from fingertip photoplethysmography (PPG) signals combined with modern deep learning (DL) techniques. Utilizing two benchmark datasets with extensive PPG recordings, we introduce ApSense, a DL model designed for OSA event onset recognition from PPG features.

ApSense presents a custom neural architecture and domain-specific feature extraction from PPG waveforms. We benchmark it against state-of-the-art (SOTA) algorithms, including RRWaveNet, PPGNetSA, AIOSA, DRIVEN, and LeNet-5. In our evaluations, ApSense demonstrated improved sensitivity, specificity, and area under the receiver operating characteristic (AUROC) on the test datasets. Furthermore, an ablation study highlighted strategic customizations of ApSense, enhancing its performance and adaptability to different datasets. ApSense demonstrates high reliability, as its outstanding results were confirmed even in high-variance datasets.

By detecting OSA events, ApSense enables the estimation of the predicted Apnea-Hypopnea Index (pAHI), which can be used for pre-screening individuals for sleep apnea in a low-cost setup. ApSense shows the potential for PPG-based OSA detection and clinical applications for pre-screening in the future.

\end{abstract}


\input{content/Introduction}
\input{content/RelatedWorks}
\input{content/DataPreparation}

\input{content/Method}

\input{content/Experiments}

\input{content/Results}

\input{content/Discussion}

\input{content/Conclusion}

\bibliographystyle{IEEEtran}
\bibliography{citelist}
\end{document}

%% file: content/Introduction.tex
\section{Introduction}
\label{sec:intro}
\IEEEPARstart{S}{leep} apnea is a common symptomatic disease caused by total (apnea) or partial (hypopnea), repeatedly interrupting respiration during sleep \cite{banno2007sleep}. Nearly one billion people worldwide suffer from sleep apnea, with certain countries having rates of over 50\% \cite{benjafield2019estimation}. Moreover, a cost-effective obstructive sleep apnea (OSA) prescreening solution is an urgent issue due to the invisible costs revealed in the study of OSA's undertreatment in many countries.


Due to their unaffordability, we must improve the existing measurement tools for massive OSA prescreening. Due to enormous demands, the Internet of Things (IoT) frameworks and devices, such as wearable sensors and data-driven AI in cloud computing, will play an essential role. Recently, one research team proposed using one type of IoT device, a wearable photoplethysmography (PPG) sensor, to replace conventional OSA prescreening tools. They reveal that the wearable PPG sensor has the potential for OSA prescreening \cite{9847226}. Also, there is scientific support for the potential of PPG in capturing the characteristics of the arterial pulse wave related to the cardio-respiratory system \cite{ALLEN2022189}, which is associated with OSA event onset.

Considering the available consumer-grade IoT devices with PPG sensors, the smartwatch is almost the only product that measures the human wrist. However, we need clinical validation that supports the usefulness of PPG measurement placed at the wrist, which is not close to the clinical setup, before encouraging smartwatch manufacturers to include OSA prescreening-related features in their commercialized products. Thus, enabling society to use wrist PPG sensors might be a substantially challenging and too time-consuming option for responding to urgent demands in massive-scale OSA recognition. In contrast to wrist PPG, fingertip PPG is more commonly used in sleep research, as presented in a few benchmark datasets \cite{10.5665/sleep.4732, LEWIS201759, bernardini2022osasud}. Conducting the feasibility study by emphasizing the performance of the OSA event onset recognition algorithm using a single fingertip PPG sensor to achieve the massive prescreening of OSA technology should be more practical and much faster. Also, some consumer-grade ring-shaped PPG sensors are available in the market. We are sure that they are ready to incorporate the findings of this paper to increase the value of their products.

Thus, this paper focuses on the fingertip PPG. We develop a novel algorithm named \textit{ApSense}, which has PPG features as the algorithm's inputs and recognizes OSA event onset as the output. We evaluate ApSense performance using the existing datasets with available related clinical data from the sleep laboratories. The outcome of this study will contribute to the potential uses of the fingertip PPG and reveal the scalability of performing OSA prescreening. More information about this work's contribution is written at the end of the next section.

The remaining parts of this paper begin with selecting the existing works as the baselines benchmarked against our proposed algorithm in Section \ref{sec:re}, followed by the contribution list of the studies. The data preparation and methodology are presented in Section \ref{sec:dat} and Section \ref{sec:med}. Experimental results are reported in Section \ref{sec:results} and discussed in Section \ref{sec:discussion}, including the future benefits from the contribution of this study. Section \ref{sec:con} finally concludes the key takeaways of this paper.

%% file: content/RelatedWorks.tex
\section{Motivation and Contribution}
\label{sec:re}
This literature summarizes previous works on developing the computational intelligent algorithm to recognize OSA events using the fingertip's OxSat or \ch{SpO2} time series values. The early study reported the advantages of using machine learning (ML), specifically an AdaBoost model built with linear discriminates (AB-LDA), over a traditional method used in the clinics, which was statistical thresholding \cite{gonzalo2019}. They exhibited the feasibility of OSA severity prediction. Later, the researchers tried to improve the performance by adding another input modality, the PPG waveform acquired from the same fingertip. The optimal sets of features extracted from OxSat and PPG waveform were proposed with optimal ML classifiers such as Fine Gaussian SVM for OSA event recognition \cite{remo2020}. Recently, the feasibility of using wrist PPG has become another alternative approach to identifying apneic events \cite{chen2021,9847226}. Although when the results are promising, due to the uncleared explanations in their articles, we cannot even reproduce the method used in their studies as a baseline for the ongoing developed algorithm -- one of the common weaknesses in the existing works related to PPG and OSA.

As an emerging research area with limited public dataset resources, deep learning (DL) approaches toward OxSat- and PPG-based OSA event recognition are currently scarce in the literature. Thus, we extend the scope of the survey to the recently developed DL-based algorithms for OSA sensing, which are biopotential-based, such as using electrocardiogram (ECG) and electroencephalography (EEG) as the sensing modalities \cite{bernardini2021aiosa, li2023deep}. Correspondingly, this scope influences how we select the existing state-of-the-art (SOTA) algorithms to benchmark against our proposed algorithm in the rest of the experiments. 

Since PPG contains derivative information of ECG reflected in cardiovascular activities, reproducing and adapting any SOTA algorithm that uses ECG to detect OSA onsets to process PPG is convincing and applicable. Those algorithms are included in the benchmark against our proposed algorithm. Meanwhile, EEG has the lowest signal-to-noise ratio among the proper non-invasive sensing on the human body. SOTA algorithms for EEG in OSA recognition are the most advanced techniques to be reproduced and applied to PPG datasets for benchmarking. In addition to priors, our research group recently published a data-driven algorithm named \textit{RRWaveNet} to estimate the respiratory rate from the PPG waveform \cite{10098530}. It is currently known as the SOTA algorithm for this specific application. As respiratory activity at night directly relates to abnormal breathing, such as OSA, we also include the modified version of RRWaveNet as another SOTA algorithm in the benchmark.

In this work, we introduce a novel \textit{ApSense} algorithm inspired by the cascading convolutional and recurrent neural network to extract spatiotemporal information concurrently, contributing to the related research fields in the following aspects.

\begin{itemize}
    \item We modify SOTA algorithms applied to other non-PPG bio-potential modalities to be compatible with PPG signal, reintroducing them under RRWaveNet, PPGNetSA, AIOSA, DRIVEN and LeNet-5 as the baselines for benchmarking against ApSense.
    \item We propose ApSense, which significantly outperforms other algorithms in high variance datasets, achieving better metrics, including the sensitivity, specificity, and the power of discrimination or the area under the receiver operating characteristic (AUROC).
    \item We present an ablation study configuring the essential components of ApSense to obtain the best AUROC. We expect the results to guide other researchers in applying and optimizing ApSense for other PPG-based applications.
\end{itemize}

\input{figure/pipeline}
\input{table/demographic}
\input{figure/AHIdist}

%% file: figure/pipeline.tex
\begin{figure}
  \center
  \includegraphics[width=\columnwidth]{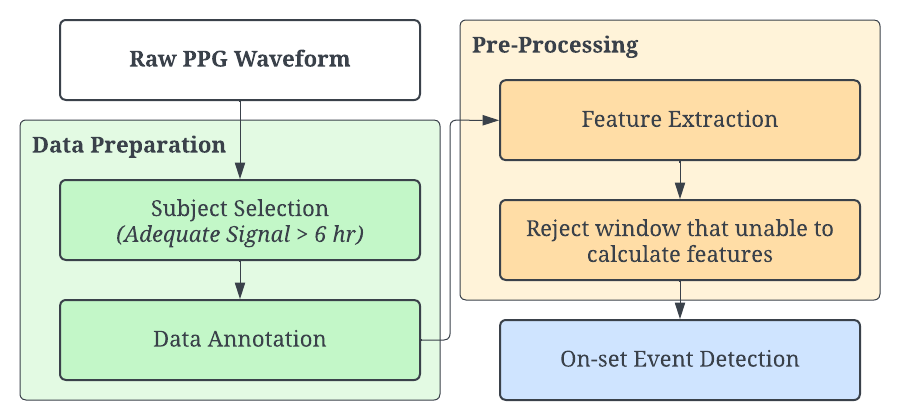}
  \caption{PPG data preparation used to develop our proposed algorithm.}
  \label{fig:pipeline}
\end{figure}

%% file: table/demographic.tex
\begin{table}[t]
\centering
\caption{Demographic Data of Participants}
\label{table:demo}

\resizebox{\columnwidth}{!}{%
    \begin{tabular}{l p{0.1\columnwidth} ll}
    \toprule[0.2em] 
    
    \textbf{Details} & & \textbf{MESA} & \textbf{HeartBEAT} \\
    
    \midrule[0.1em]
    \# of subjects & & 276 & 211 \\
    Age (years) & & 67.95 $\pm$ 8.49 & 62.73 $\pm$ 7.08 \\
    F/M ratio & & 0.97 & 0.44 \\
    BMI (kg/m\textsuperscript{2}) & & N/A & 34.13 $\pm$ 6.18 \\
    AHI (events/hr) & & 21.82 $\pm$ 17.79 & 25.03 $\pm$ 8.36 \\

    \midrule[0.1em]

    PPG Sampling rate (Hz) & & 256 & 75 \\
    \ch{SpO2} Sampling rate (Hz) & & 1 & 3 \\
    
    \midrule[0.1em]
    
    Total windows & & 356,480 & 523,621 \\
    Rejected windows & & 32,253 & 39,039 \\
    Rejected rate & & 7.46\% & 9.05\% \\
    Remaining windows & & 324,227 & 484,582 \\
    \# of windows for training & & 232,333 & 344,684 \\
    \# of windows for validate & & 58,084 & 86,172 \\
    \# of windows for test & & 33,810 & 53,726 \\
    
    \bottomrule[0.2em]   
    \end{tabular}
}

\vspace{0.5em}
\scriptsize{{\raggedright \textbf{Abbreviations:}
F/M; Females/Males,
BMI; Body Mass Index,
AHI; Apnea-Hypopnea Index \par}}

\end{table}

%% file: figure/AHIdist.tex
\begin{figure}[t]
    \centering
    \includegraphics[width=0.48\columnwidth]{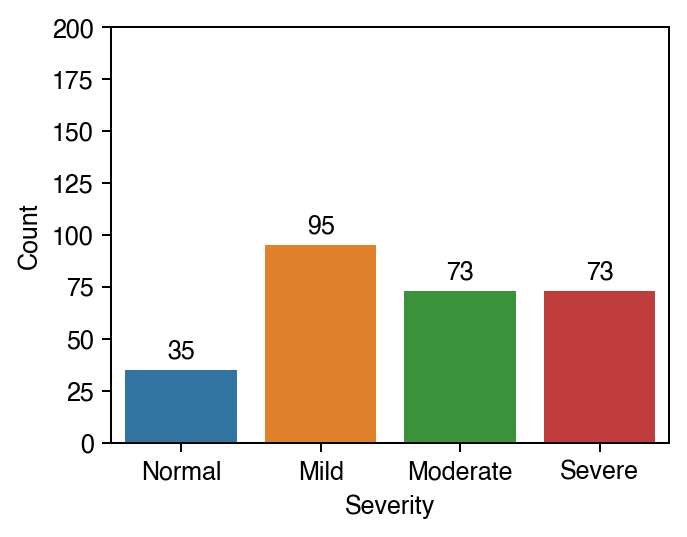}
    \includegraphics[width=0.48\columnwidth]{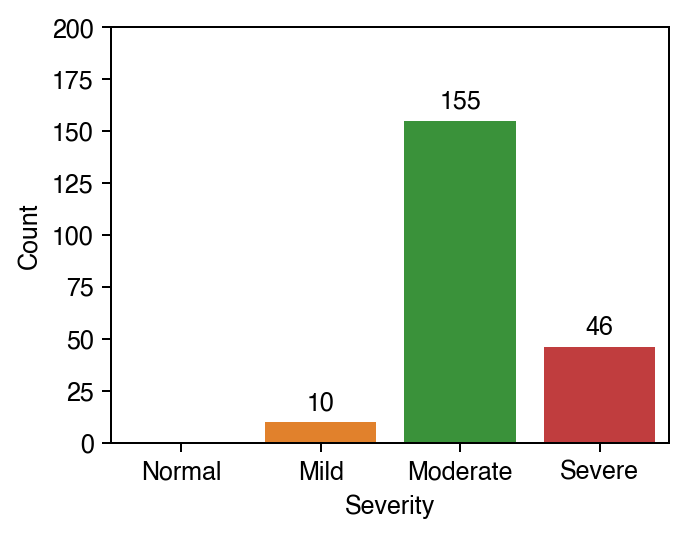}
    \caption{AHI distributions of subjects from MESA (left) and HeartBEAT dataset (right) classified by severity \cite{10.1093/sleep/22.5.667}.}  
    \label{fig:AHI}
\end{figure}

%% file: content/DataPreparation.tex
\section{Data Preparation}
\label{sec:dat}
Figure \ref{fig:pipeline} illustrates the pipeline of PPG datasets preparation for algorithm development. The rest of this section follows the data preparation, segmentation, annotation, and feature extraction.

\subsection{Datasets}
Two widely used PSG datasets in this study are the Multi-Ethnic Study of Atherosclerosis (MESA) \cite{10.5665/sleep.4732} and Heart Biomarker Evaluation in Apnea Treatment (HeartBEAT) \cite{LEWIS201759}. Both consist of PPG signals in red color wavelength, which is in the interest of our study \cite{10.1093/jamia/ocy064}. 

MESA is a six-center longitudinal collaborative project supported by the NHLBI that examined factors related to the development of subclinical cardiovascular disease (CVD) and the progression of subclinical CVD to CVD. In this work, we utilize MESA version 0.3.0, which consists of sleep exams conducted between 2010 and 2012 and includes full overnight unattended PSG.

HeartBEAT is a multi-center Phase II randomized controlled trial that evaluated the effects of Positive Airway Pressure (PAP) therapy over a three-month intervention period in patients with CVD or CVD risk factors and moderate to severe OSA (AHI ranged from 15 to 50). The study's primary outcome was a 24-hour blood pressure profile.

To control the experiment according to the scope of work, we only consider people with good PSG signal quality graded by the publisher in this study. In the case of MESA, we include subjects whose quality of all signals is good for over six hours of the entire sleep time or equivalently assessed with an \emph{outstanding} grade (score $= 7$) according to its data documentation guide. While HeartBEAT contains much less amount of samples, we choose subjects whose signal quality is good for at least four hours or equivalently graded as at least \emph{excellent} grade (score $\geq5$). Table \ref{table:demo} summarizes the patients' demographics from both datasets, including the apnea-hypopnea index (AHI). Figure \ref{fig:AHI} shows the distributions of AHI from the samples used in this study.

\input{figure/Annotation}
\input{figure/Pulse}
\subsection{PPG-based Apnea Event Annotation}

Figure \ref{fig:label} presents the characteristics of the signals from three sensing modalities: the nasal airflow, the pulse oximeter, and PPG. The nasal airflow is vital in recognizing breathing abnormalities, including OSA (apnea event). Pausing breathing noticed via the airflow often comes with the dropping of \ch{SpO2} measured via the pulse oximeter or so-called oxygen desaturation. As reported in \cite{dumitrache2013role}, there is a lag between breathing pause and \ch{SpO2} drop. Since this paper aims to demonstrate the potential applications of PPG in recognizing OSA, we focus on PPG characteristics whose waveforms appear in the bottom row of Figure \ref{fig:label}. Since the decrease in pulse wave amplitude (PWA) in PPG results from sympathetic nervous system arousal caused by the lowered \ch{SpO2} level during OSA, we identify any PWA drop that coincides with the oxygen desaturation and lags to the airflow event. We propose the following procedure for annotating PPG windows with OSA events.

First, we use the sliding window approach to segment the PPG signal into 60-second periods with 50\% overlapping. Then, we identify the timestamp at the end of each apnea or hypopnea period that the sleep specialist annotated while examining through the complete PSG system or multiple sensing modalities. Suppose the identified timestamp is inside the first half of a PPG window; we mark it as an OSA event, otherwise non-event or normal. As mentioned earlier, this procedure ensures that any interested PPG window includes informative features such as PWA, which is helpful for OSA recognition. The OSA event marking is similar to what was introduced in the previous work, confirmed by PPG and \ch{SpO2} \cite{remo2020}.

\subsection{Feature Extraction}
\label{Feature_Extraction}

Sleep apnea can result in intermittent breathing pauses, decreasing oxygen intake, and causing stress on the cardiovascular system. Hence, retrieving features related to the cardiovascular circulatory from the PPG waveform is crucial. According to Figure \ref{fig:pipeline}, we extract seven important features before passing them to the OSA event onset recognition algorithm. We first filter the raw PPG signal in each window using the second-order Chebyshev type II band-pass filter (0.1 - 20 Hz) using the Scipy library \cite{2020SciPy-NMeth}. Then, we use the moving average of 64 time points to filter out high-frequency noise. As the feature calculation involves peak-to-peak changes illustrated in Figure \ref{fig:pulse}, we denote each pulse by obtaining the top and bottom peaks of the waveform by identifying local maxima and minima and ensuring that they are at least 30 time points apart (by using the scipy.agrelextrema function). Then, we calculate features from all $N$ pulse waveforms in each input PPG window as follows:
\\
\\
\textbf{Pulse wave amplitude (PWA):} Pulse wave amplitude (PWA) refers to the amplitude of the PPG pulse from the pulse's bottom peak to the top peak. We calculate PWA as follows:
\begin{equation}
\label{pwa_fnc}
\text{PWA}_{n} = f(t_{tp,n})-f(t_{bp,n})
\end{equation}

where $f(t)$ refers to the amplitude of PPG at time $t$. $t_{tp,n}$ and $t_{bp,n}$ are the time when a PPG pulse $n$ is at peak and trough, as demonstrated in the first pulse of Figure \ref{fig:pulse}. PWA directly correlates with finger blood flow. Drops in PWA are strongly associated with changes in EEG measurements, suggesting that they are an appropriate proxy for changes in cortical activity \cite{delessert2010pulse}. During an apnea event, the PWA increases until it reaches its peak and decreases during the hyperventilation period \cite{GROTE2003141}.
\\
\\
\textbf{PP interval (PPI):} PP interval (PPI) is the interval between the top peaks between two consecutive pulses:
\begin{equation}
\label{ppi_fnc}
\text{PPI}_{n} = t_{tp,n+1}-t_{tp,n}
\end{equation}

PPI can be used to determine the pulse rate. Pulse rate variability (PRV) has been used as a surrogate of heart rate variability (HRV) \cite{CYGANKIEWICZ2013379}, which can evaluate automatic nervous system activity in clinical practice. It also has been used in screening algorithms of OSA \cite{Nakayama_2019, KHANDOKER2011204}.
\\
\\
\textbf{Derivative of PWA (dPWA):} The derivative of PWAs is the difference between two consecutive PWAs of the PPG signal. Assuming that $\text{PWA}_{n}$ is the PWA of pulse $n$,
\begin{equation}
\label{dpwa_fnc}
\text{dPWA}_{n} = \text{PWA}_{n+1}-\text{PWA}_{n}
\end{equation}
\textbf{Derivative of PPI (dPPI):} Similar to the dPWA, it is the difference between two consecutive PPIs. Assuming that $\text{PPI}_{n}$ is the PPI between pulse $n+1$ and pulse $n$,
\begin{equation}
\label{dppi_fnc}
\text{dPPI}_{n} = \text{PPI}_{n+1}-\text{PPI}_{n}
\end{equation}
\textbf{Systolic phase duration (SPD):} Systolic phase is a phase in the cardiac cycle when the heart contracts. The phase duration is defined by the time from the trough to the following peak. SPD is defined as
\begin{equation}
\label{spd_fnc}
\text{SPD}_{n} = t_{tp,n}-t_{bp,n}
\end{equation}
\textbf{Diastolic phase duration (DPD):} Diastolic phase is another phase in the cardiac cycle that occurs as the heart relaxes. This phase duration is defined by the time from the peak to the next trough. Therefore,
\begin{equation}
\label{dpd_fnc}
\text{DPD}_{n} = t_{bp,n+1}-t_{tp,n}
\end{equation}

In most people, blood pressure drops 10\%-15\% from its diurnal value during sleep. Patients who do not experience this drop in blood pressure are called non-dippers. The prevalence of OSA severity is associated with the prevalence of non-dipping, especially in patients with OSA compared to those without \cite{10.1016/S0895-7061(01)02143-4, Suzuki1996-uf}. It could be postulated, therefore, that OSA may affect the measurements of systolic and diastolic blood pressure.
\\
\\
\textbf{Pulse area (PA):} Pulse area is calculated by integrating the PPG pulse from the start of the systolic phase until the end of the diastolic phase.
\begin{equation}
\label{pa_fnc}
\text{PA}_{n} = \sum_{i=t_{bp,n}}^{t_{bp,n+1}} f(t_{i})
\end{equation}

Since the number of pulse waveforms, $N$, differs across the windows, we resample to a 60-point input feature vector. All extracted features are further normalized channel-wise.

Table \ref{table:demo} reports the number of extracted 60-point windows. When a window has low signal quality, the feature extraction may fail. This contributes to the rejected rate, accounting for $7.46\%$ and $9.05\%$ in MESA and HeartBEAT datasets. The remaining windows are split into training and test sets, explicitly detailed in Section \ref{sec:perf_eval}.

\input{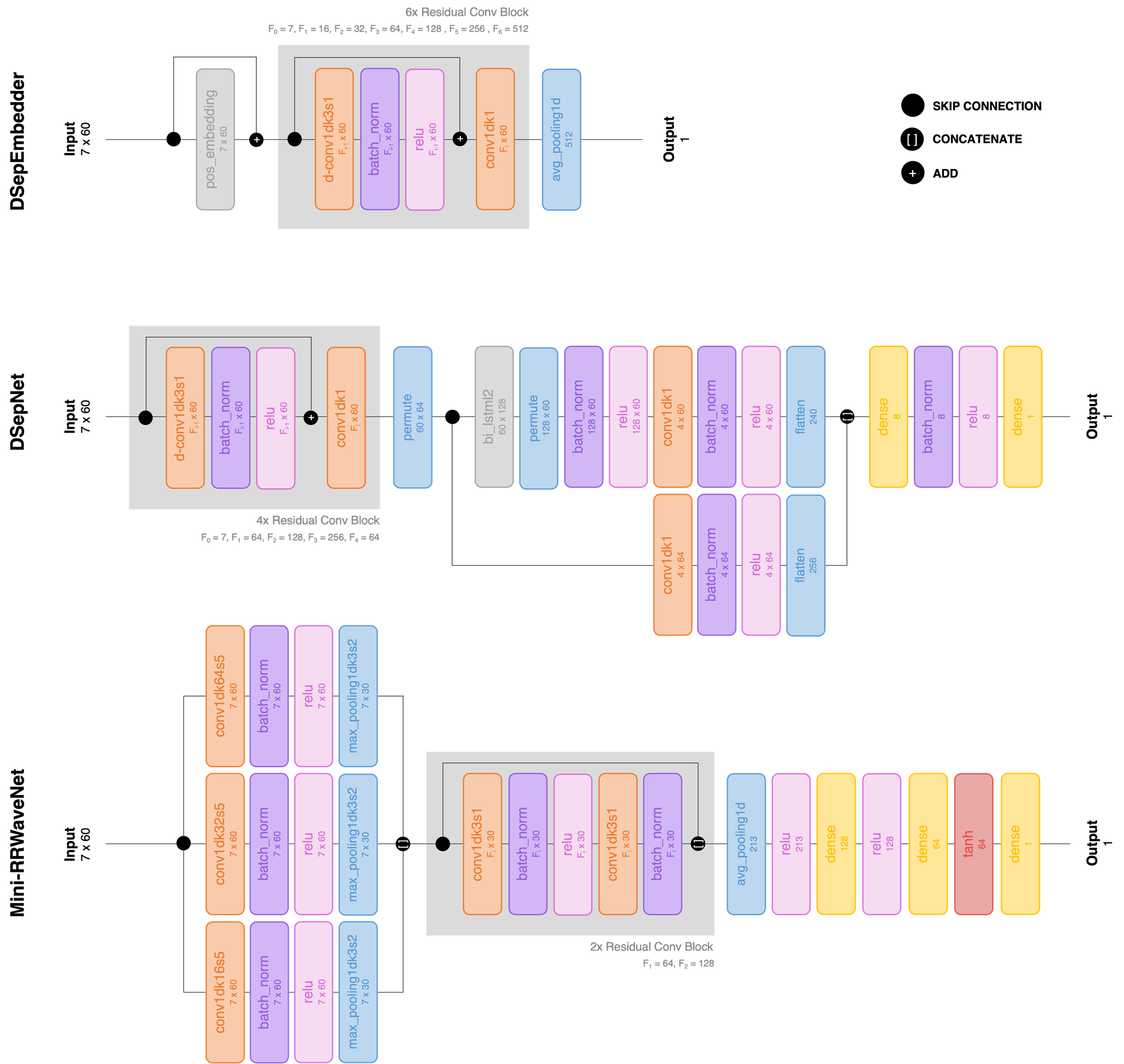}
\input{figure/dmodel_2}

%% file: figure/Annotation.tex
\begin{figure}
  \center
  \includegraphics[width=\columnwidth]{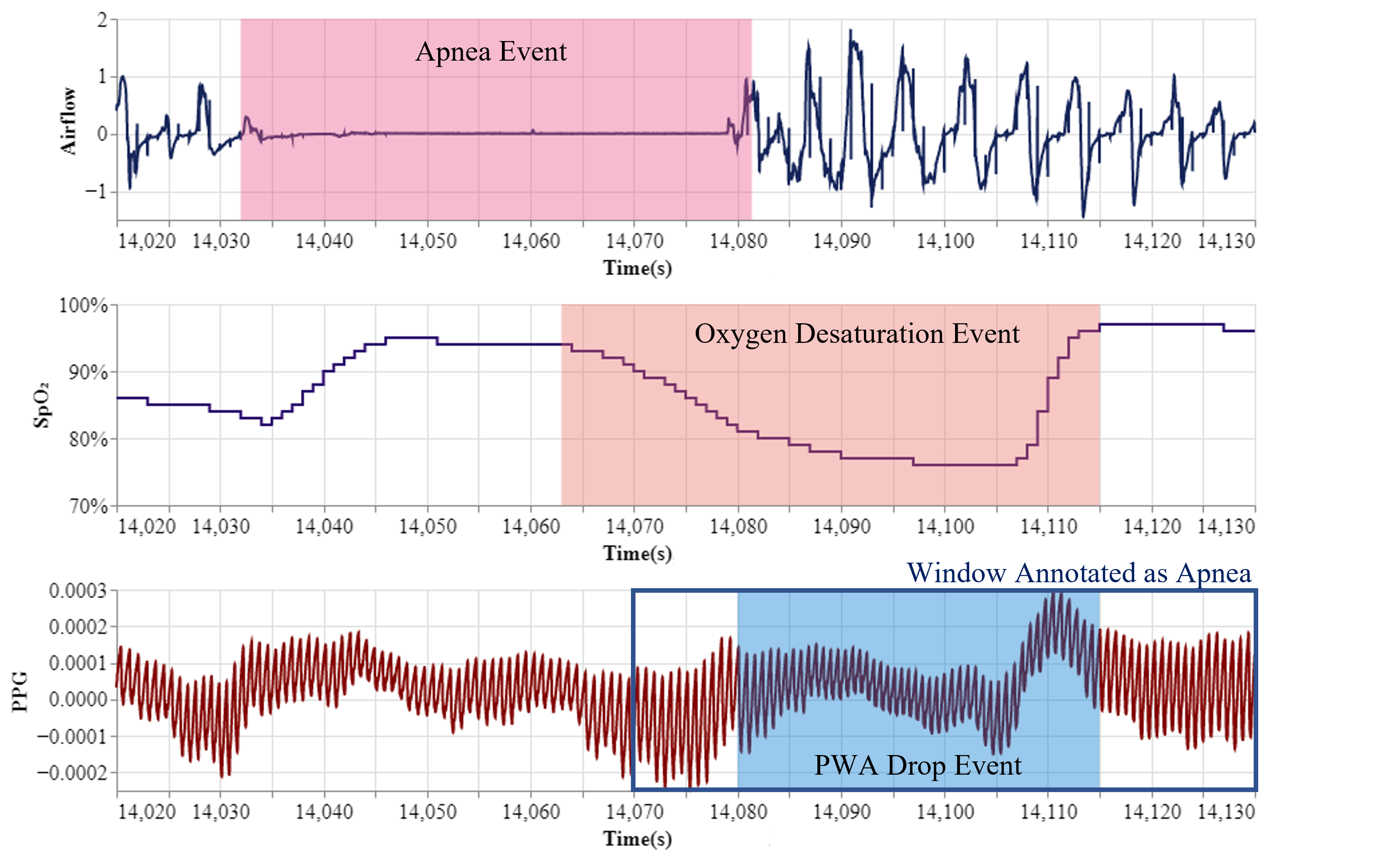}
  \caption{Example of the signal from air flow, \ch{SpO2}, and PPG sensors.}
  \label{fig:label}
\end{figure}

%% file: figure/Pulse.tex
\begin{figure}
  \center
  \includegraphics[width=0.7\columnwidth]{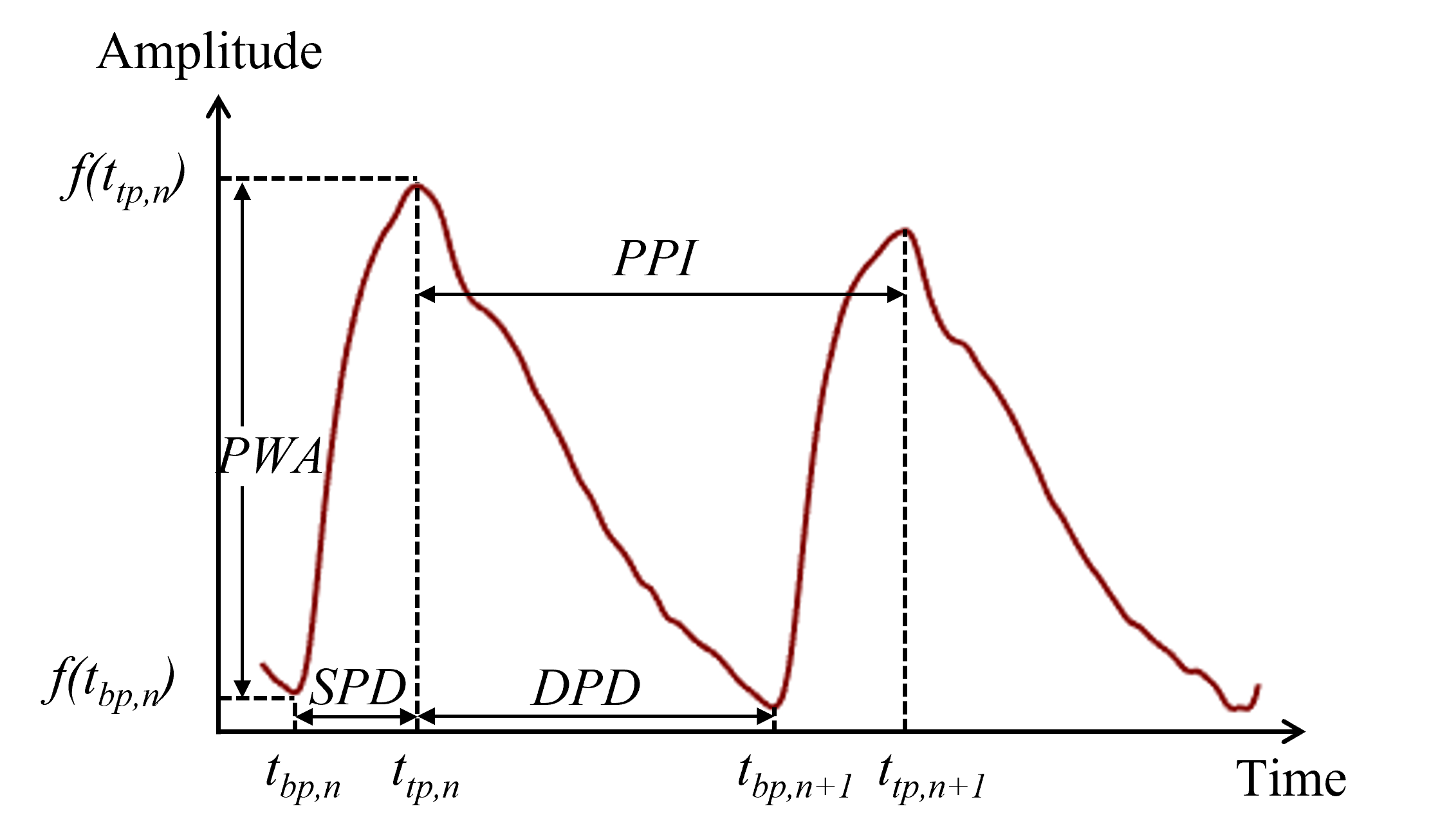}
  \caption{Pulse waveform of PPG AC part with pulse wave characteristics.}
  \label{fig:pulse}
\end{figure}

%% file: figure/dmodel.tex
\begin{figure*}
  \center
  \includegraphics[scale=1.00,width=2\columnwidth]{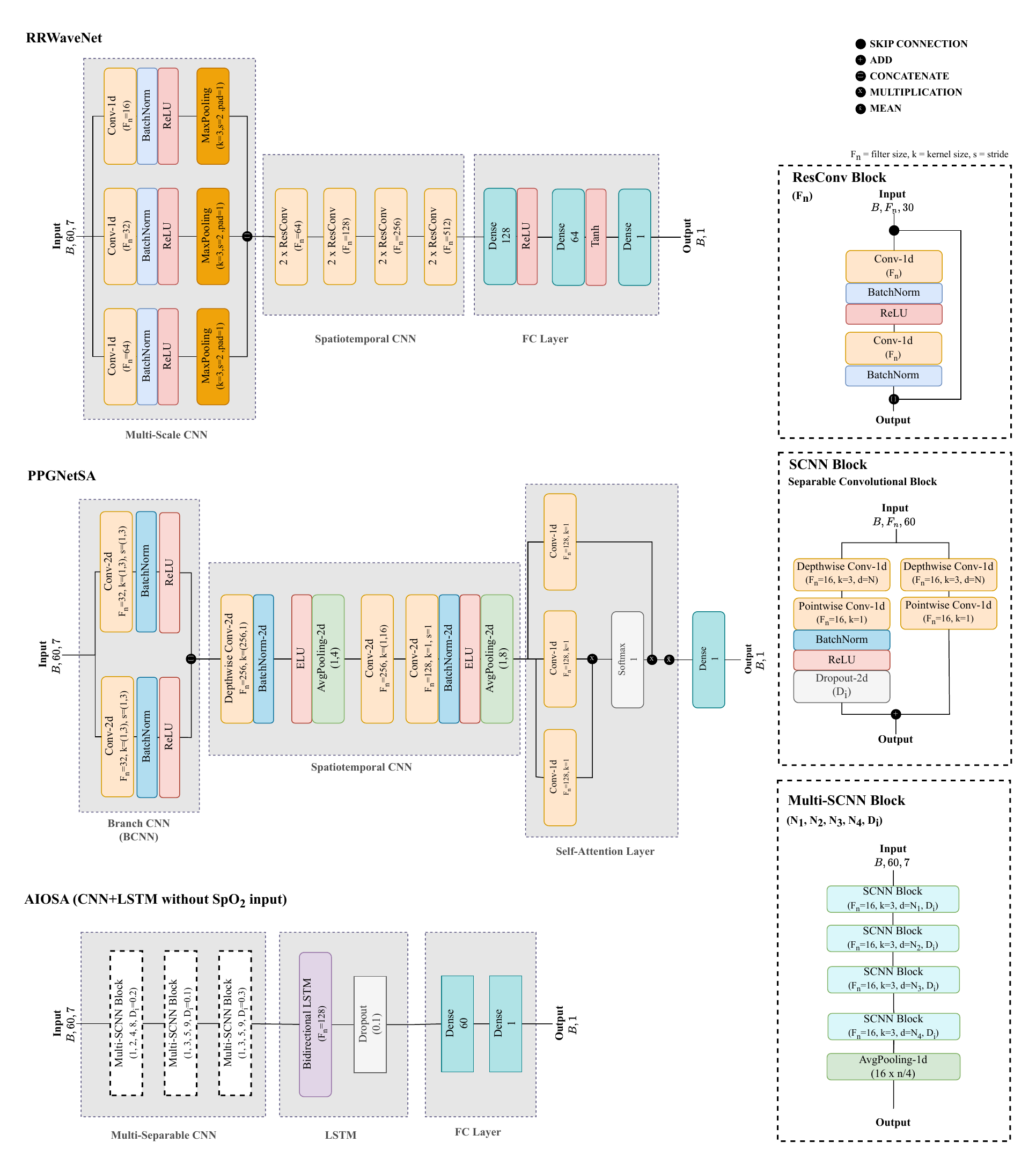}
  \caption{Deep neural network structures of the baseline OSA event onset detection algorithms (RRWaveNet, PPGNetSA, and AIOSA)}
  \label{fig:model}
\end{figure*}

%% file: figure/dmodel_2.tex
\begin{figure*}
  \center
  \includegraphics[scale=1.00,width=2\columnwidth]{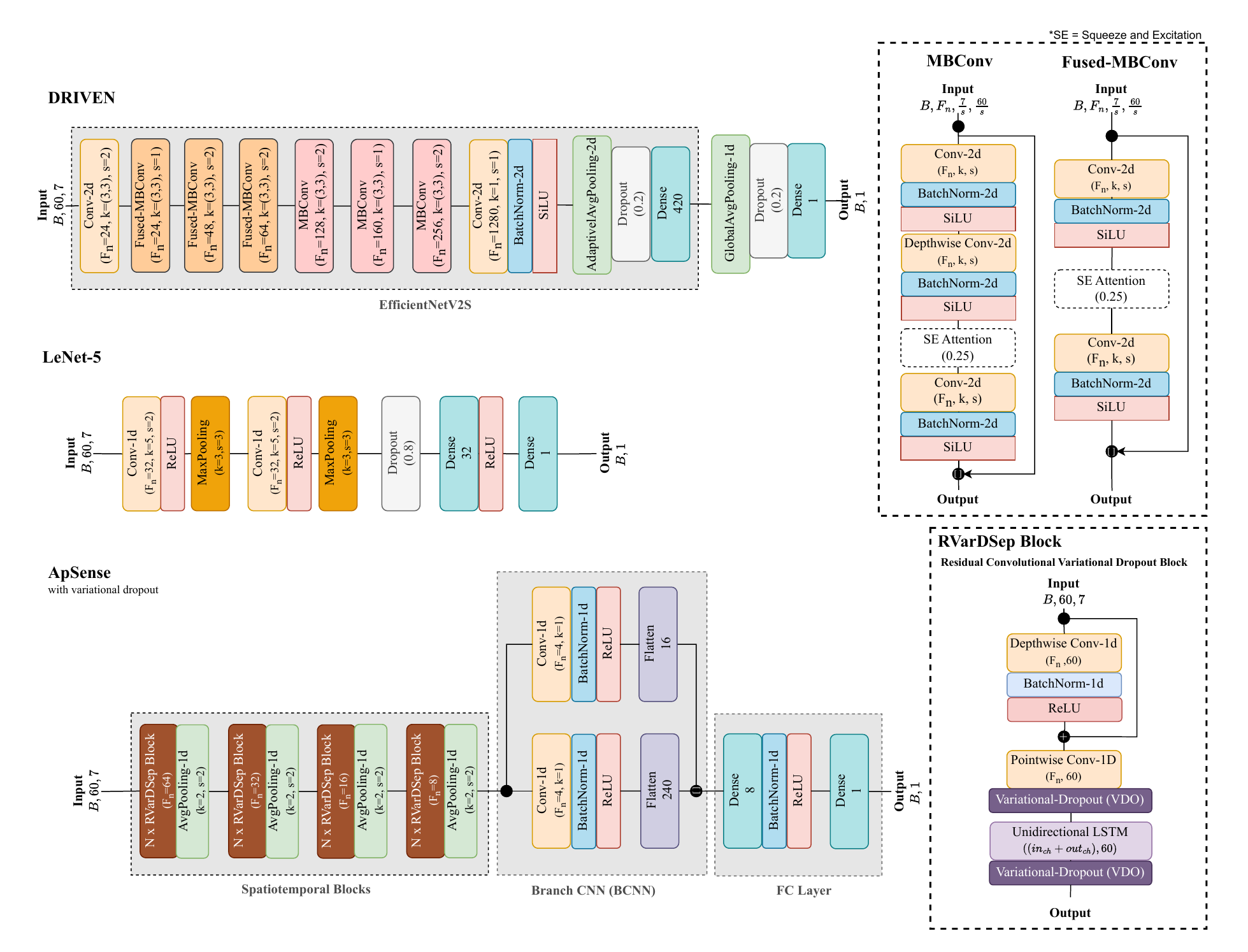}
  \caption{Deep neural network structures of the baseline OSA event onset detection algorithms (DRIVEN, LeNet5) and proposed algorithm (ApSense)}
  \label{fig:model2}
\end{figure*}

%% file: content/Method.tex
\section{Method}
\label{sec:med}
Here, we explain deep neural network structures, which are the backbone of the proposed (ApSense) and SOTA algorithms of OSA event onset recognition (RRWaveNet, PPGNetSA, AIOSA, DRIVEN, and LeNet-5) as illustrated in \autoref{fig:model} and \autoref{fig:model2}. SOTA algorithms are modified from their original version to be compatible with PPG datasets and to ensure fair benchmarking against ApSense. All algorithms start with the input format of ($B$, 60, 7), indicating the number of batch size ($B$), time points, and PPG waveform features, respectively. This section mainly clarifies each neural network component for reproducibility. We also publish the source codes of these algorithms on GitHub.\footnote{Repository: \url{https://github.com/IoBT-VISTEC/ApSense} will be publicly accessible after the acceptance of the paper}.

\subsection{ApSense}
ApSense comprises three essential components: (i) multiple spatiotemporal blocks, (ii) a branch convolutional neural network (BCNN), and (iii) the fully connected layers, as illustrated in \autoref{fig:model2}.

\subsubsection{Multiple spatiotemporal blocks}
We introduce multiple spatiotemporal blocks inspired by previous work utilizing CNN-BiGRU with stacked spatiotemporal blocks for OSA recognition using ECG time-series data. Their method can extract local spatial features of the R-peaks and global temporal features of the heartbeat intervals, outperforming a simple combination of CNN and LSTM networks. Furthermore, they highlighted that a single spatiotemporal block model cannot effectively extract high-level spatiotemporal information and used up to four blocks to learn high-level feature information \cite{chen2022spatio}.

On the other hand, our proposed spatiotemporal block consists of a series of RVarDSepBlocks stacked arbitrarily. Each RVarDSepBlock is defined by a fixed sequence of different filter numbers ($F_{n}$ = 64, 32, 16, and 8), repeated $M$ times. Each RVarDSepBlock includes a residual depthwise separable 1D-convolution splitting the computation into two steps: depthwise and point-wise convolutions, followed by a batch normalization, a ReLU activation, and a unidirectional LSTM module. The LSTM module incorporates a variational dropout layer, masking input dropout with a probability of 0.1 to avoid overfitting in a sequence-based model \cite{benjafield2019estimation}. The output channel in this layer is updated to the sum of the input (in\_ch) and output (out\_ch) channels. After each RVarDSepBlock, an average pooling operation is applied to reduce size. Subsequently, the final production of RVarDSepBlock is fed into two parallel convolutional branches.

\subsubsection{Branch Convolutional Neural}
Taking inspiration from multi-scale convolution techniques \cite{szegedy2017inception,shi2023exploiting}, we construct a two-branch CNN (BCNN) architecture to capture features at different resolutions ideal for handling output from the spatiotemporal blocks (RVarDSepBlocks). The output from RVarDSepBlocks is processed through two separate CNN branches. The upper branch employs a 1 $\times$ one convolution for channel-wise operations, reducing the data dimensionality and capturing global relationships across channels. In contrast, the lower branch utilizes regular 1D convolutions to capture local patterns within each time point. These branch outputs are concatenated and flattened before passing into two fully connected layers.

\subsubsection{Dense Fully Connected Layers}
We employ two fully connected layers. The first accepts the flattened and concatenated outputs from both branches. Following this, batch normalization and ReLU activation transform data into a lower-dimension space (output size 8). The second layer generates a prediction by leveraging ReLU's non-linearity for richer data representation. This output layer condenses the features into a final output distinguishing OSA onset from non-OSA events.

\subsection{RRWaveNet}
RRWaveNet \cite{10098530} is a recent SOTA method that estimates respiratory rate using raw PPG signal as an input. We adapted it to predict apnea event recognition.

As illustrated in \autoref{fig:model}, RRWaveNet starts with a Multi-Scale CNN block containing three parallel multi-scale convolutional layers, with filter sizes ($F_{n}$) of 16, 32, and 64 - each followed by batch normalization, ReLU activation and the max-pooling layer. Concatenating the results from all three branches produces the final output.

The spatiotemporal CNN block consists of eight ResConv blocks with different filter sizes ($F_{n}$ = 64, 128, 256, and 512), repeated twice before the following filter size. A ResConv block consists of two residual convolutional blocks arranged sequentially. Each block includes a convolutional layer, batch normalization, ReLU activation, another convolutional layer, and final batch normalization. A skip connection concatenates the input and output of the convolutions.

The final module, composed of fully connected layers, starts with ReLU activation and a fully connected layer, followed by a $\tanh$ activation and a final dense layer that produces the OSA prediction. The number of features in the fully connected layers progressively decreases to 128, 64, and finally 1, resulting in a single value prediction for apnea event detection.


\subsection{PPGNetSA}
The PPGNetSA model is constructed based on the Res-Net architecture \cite{li2023deep}, initially devised for apnea detection using raw EEG data. Res-Net draws inspiration from transformers, utilizing a self-attention mechanism instead of fully connected layers while integrating branch CNNs and spatiotemporal blocks. However, we slightly adapt Res-Net to accommodate PPG waveform inputs, namely the PPGNetSA model.

As illustrated in \autoref{fig:model}, the PPGNetSA model consists of three customized blocks. The first block, the branch block convolutional layer (BCNN), employs two parallel convolutional layers with filter size ($F_{n}$) of 64 and stride ($s$) of 3. This BCNN utilizes two kernel sizes ($k$): one equal to half the window size and the other with $k$ = 3. Following this, batch normalization and ReLU activation are independently applied to the outputs of these layers before they are concatenated. This strategic approach exploits the larger kernel size convolution for extracting frequency features and the smaller kernel size for capturing transient features.

The second block, known as the spatiotemporal convolutional block, is responsible for extracting specific spatial and temporal information through the use of batch normalization, average pooling, and ELU activation on a depthwise convolutional layer with $F_{n}$ and $k$ of 256. Following this, two separable convolution layers capture temporal features with $F_{n}$ of 256 and 128 for each.

Finally, the self-attention block is applied. The output from the spatiotemporal block is fed into three convolutional layers ($F_{n}$ = 128 and $k$ = 1) to obtain three sets of vectors. The first two vectors undergo dot multiplication, divided by the square root of the feature map dimension, and are then processed through a softmax function. The resulting output is multiplied by the last vector to produce the self-attention output, which is utilized as an input for a global average pooling layer and passed through a fully connected layer to generate the final prediction output.

\subsection{AIOSA}
AIOSA is a deep learning model architecture initially designed for ECG-based OSA (obstructive sleep apnea) onset event recognition \cite{benjafield2019estimation}. AIOSA employs a CNN-LSTM neural network architecture, leveraging depth-wise separable convolutions to reduce computational complexity. This approach includes using various dilation rates to capture different temporal frames of the input data, followed by an LSTM module. For a fair comparison with other baseline models, the proposed unimodal variant of AIOSA (utilizing ECG without \ch{SpO2} input) is adapted to use single PPG inputs instead. This adaptation ensures a consistent basis for evaluating the performance of all models.

The unimodal variant of AIOSA consists of (i) three multiple separable CNN blocks (Multi-SCNN block) instead of the original four blocks due to the smaller input shape, (ii) an LSTM module, and (iii) fully connected layers. Each Multi-SCNN block has four SCNN blocks arranged in a series. A SCNN block comprises a residual depthwise separable 1D-convolution (incorporating depthwise and pointwise convolutions) with dilation, followed by batch normalization, ReLU activation, and spatial dropout. Within each SCNN block, the number of filters ($F_{n}$ = 16) and the kernel size ($k$ = 3) are constant, while the dilation ($d$) and the spatial dropout ($D$) vary within and between the SCNN block. The average pooling operation attaches at the end of each Multi-SCNN block. The Multi-SCNN block output then enters a bidirectional LSTM module, and a dropout is applied. The output of this module is passed to two fully connected layers. Designed initially to predict OSA per time point with a final 60-unit dense layer size, we adapt the size of the final dense layer to 1 for OSA prediction per window.


\subsection{DRIVEN}
DRIVEN is the current state-of-the-art home-based sleep apnea estimation designed to leverage multiple physiological signals for accurate detection \cite{retamales2024towards}. The adapted version of the DRIVEN architecture builds upon this advanced framework, specifically tailored to handle multidimensional inputs. Unlike the original model, which used three EfficientNetV2S blocks to process unidimensional data from three separate channels, the adapted version now employs a single EfficientNetV2S block to handle a 7-channel PPG feature set. This modification simplifies the architecture while retaining the deep convolutional network's capability to process complex multidimensional data.

As illustrated in \autoref{fig:model2}, the DRIVEN model comprises four customized blocks: (i) a stem convolution layer, (ii) multiple stages of Mobile Inverted Bottleneck Convolution layers (MBConv) with squeeze-and-excitation modules, (iii) an adaptive average pooling layer, and (iv) a fully connected layer for classification output. Each MBConv block includes a 2D convolutional layer (Conv-2d), batch normalization (BatchNorm-2d), SiLU activation, depthwise convolution (Depthwise Conv-2d), and SE attention mechanisms. The Fused-MBConv block integrates these elements with a direct SE attention mechanism following the initial Conv-2d and BatchNorm-2d layers. After these MBConv stages, an adaptive average pooling layer reduces the spatial dimensions, preparing the features for the final classification. The model extends from the EfficientNetV2S by adding global average pooling of the pooled features, then passing through a fully connected layer to produce the classification output.

The DRIVEN model configuration employs specific settings for its convolutional layers, ensuring efficient feature extraction and robust performance detecting sleep apnea events from PPG signals. The architecture utilizes kernel sizes ($k$) of 3 across all layers, with strides ($s$) set to [1, 2, 2, 2, 1, 2] to manage the downsampling of spatial dimensions at different stages. The expand ratios are configured as [1, 4, 4, 4, 6, 6], which help adjust the layers' width to capture more complex patterns. The number of filters ($F_n$) is set to [24, 48, 64, 128, 160, 256], ensuring a progressive increase to handle varying levels of feature complexity. Additionally, squeeze-and-excitation (SE) ratios of 0.25 are applied in later stages to enhance feature recalibration and a dropout rate of 0.2 to prevent overfitting.

\subsection{LeNet-5}
The adjusted model of LeNet-5 adapts from the original ECG-LeNet-5 architecture \cite{wang2019sleep}. Initially designed for sleep apnea detection on single-lead ECG signals, LeNet-5 consists of two convolutional, pooling, and fully connected layers. It demonstrates its robustness and efficiency in extracting relevant features from physiological data.

In our adaptation, the model inputs are PPG features instead of the original ECG signals while preserving other parameters from the original LeNet-5 configuration. The motivation behind this adaptation stems from the need to leverage PPG signals while maintaining the effectiveness of the LeNet-5 architecture in detecting apnea events.

As depicted in \autoref{fig:model2}, the LeNet-5 model starts with the input layer that takes seven-dimensional features extracted from PPG data. The first convolutional layer applies 32 filters with a kernel size of 5 and a stride of 2, followed by a max pooling layer with a pooling size of 3 and a stride of 3. The second convolutional layer has 64 filters of the same stride and kernel size, again followed by the same max pooling layer. To address overfitting, a dropout layer with a dropout rate of 0.8 is inserted between the final convolutional and the fully connected layer. The fully connected layer reduces the output to 32 nodes, utilizing ReLU activation functions.

%% file: content/Experiments.tex
\section{Experimental Setup}
In this section, we assess the capability of ApSense through three experiments. Initially, in Experiment I, we optimize ApSense's configuration to identify the optimal performance for comparison with the SOTA algorithms in Experiment II. Following these, we evaluate the feasibility of using ApSense to estimate AHI overnight in Experiment III. Each experiment addresses the same handling of imbalanced data, network training, and network testing, which explain the following contents.

\subsection{Handling Imbalanced Data}

\begin{figure}
  \center
  \includegraphics[width=1.05\columnwidth]{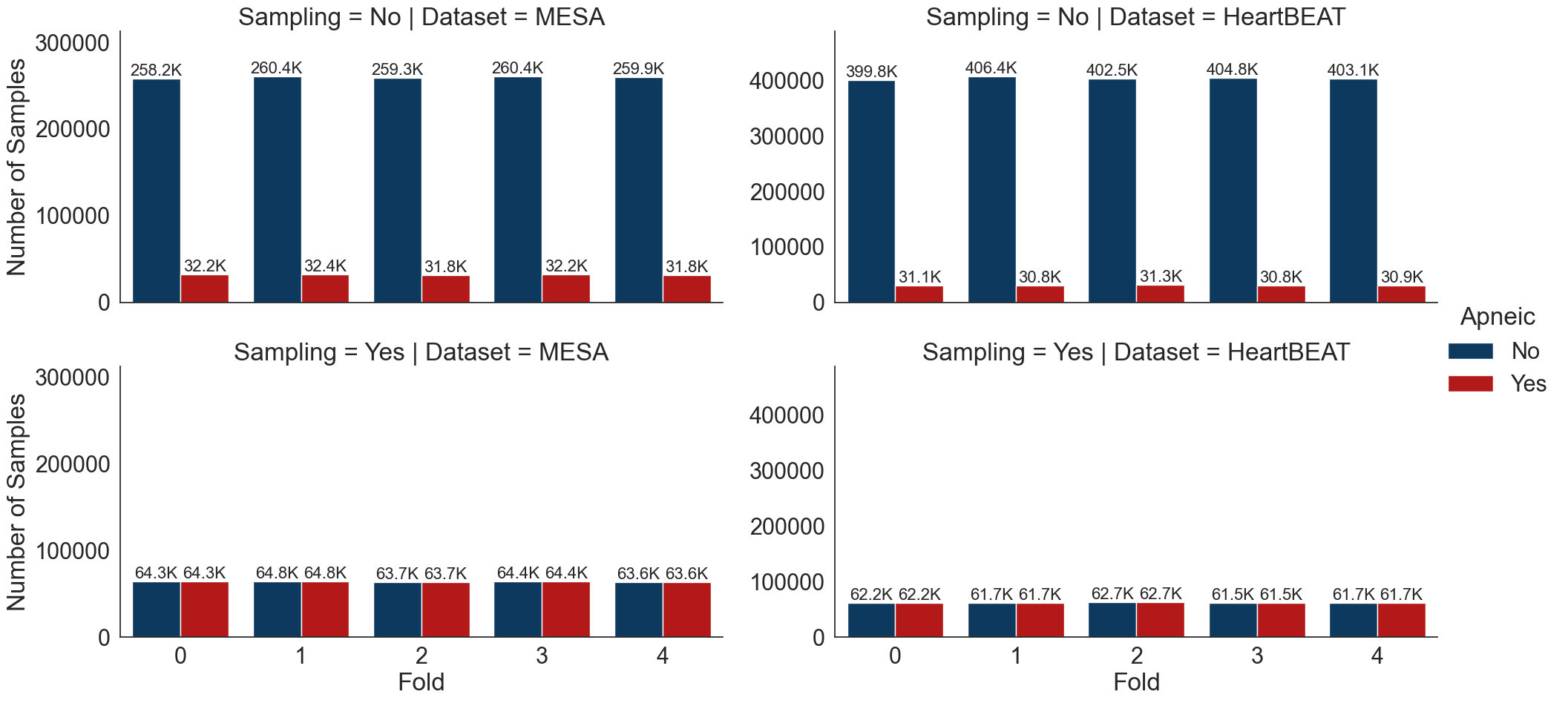}
  \caption{Number of sample labels after applying hybrid sampling method using undersampling and oversampling to handle imbalance dataset from two datasets and across the five folds. The label distribution becomes uniform after applying the technique.}
  \label{fig:Hybrid Method}
\end{figure}

Illustrated in \autoref{fig:Hybrid Method} (top), prior to employing the sampling method, the number of non-apneic event windows (blue) in the MESA and HeartBeat datasets was approximately 8 and 13 times greater than that of apneic events (red), respectively. To mitigate bias towards the non-apneic class, we apply a hybrid method for handling imbalanced data by combining oversampling and undersampling techniques \cite{choirunnisa2018hybrid}. First, we oversample the apneic class to increase the number of minority class instances by 100\%. For example, if a minority class contains 100 samples, 100 additional samples are generated, resulting in 200 minority-class samples. Consequently, we perform random undersampling to remove non-apneic samples (the majority class) to match the number of minority class instances in each iteration before entering the training process, as presented in \autoref{fig:Hybrid Method} (bottom).

\subsection{Network Training}
All algorithms in this study are trained for 1000 epochs utilizing the Adam optimizer, incorporating a learning rate of 0.001 and a batch size of 1024. Binary cross entropy is the chosen loss function. An early stopping technique is employed to mitigate potential overfitting inherent in deep neural networks, with patience of 30 epochs determined by the validation loss. The implementation of the proposed network is carried out in Python 3.6.9 using PyTorch. The experiment is executed on NVIDIA Tesla V100 SXM2 1 TB computational hardware.

\subsection{Network Validation and Testing}\label{sec:perf_eval}
In this study, our focus lies on subject-independent validation to showcase the practical applicability of our approach. Since the system has not previously encountered data from new patients, we simulate this scenario by segregating test subjects from training subjects. We underscore the importance of evaluating the proposed methods or algorithms in the OSA event recognition task.

For each dataset, we generate five development sets (training combined with validation sets) and test sets with the number of subjects at an 80:20 ratio, respectively. The same prepared sets are used to evaluate all algorithms for fair comparison, including in the benchmarking. Each subject's samples must belong to only one among the development and the test set for subject independence. Moreover, we ensure that the test set balances subjects with AHI $\le$ 15 and AHI $>$ 15 to reduce the bias evaluation affected by OSA's severity.

In the first step of the algorithm optimization, we shuffle all the subjects' windows within each development set and then conduct five-fold cross-validation. Then, the best model from the development sets is selected to evaluate its performance on the test set. In summary, five inner stratified folds are among the development sets for training and validation. Then, five outer folds are stratified between the inner folds and the test sets. We call this evaluation scheme \emph{nested five-fold cross-validation}. As we generate five test sets, we can produce the corresponding evaluation metrics, including accuracy, macro F1 score, sensitivity, specificity, and area under the receiver operating characteristics (AUROC) per model per dataset. Then, the five values are aggregated and reported as an average percentage with standard deviation.

\subsection{Experiment I: ApSense Configuration}
This experiment aims to determine the optimal ApSense configuration. The default ApSense consists of three primary components: spatiotemporal blocks created from the cascade of RVarDsepBlock with various filter sizes, the branch CNN (BCNN), and the FC layers. We can optimize ApSense performance by adjusting $M$, VDO, and BCNN, which are hyper-parameters representing the number of stacked RVarDsepBlock for each filter size, LSTM variational dropout, and branch CNN, respectively.

The assumptions for adjusting these parameters are as follows. An increasing number of spatiotemporal blocks affects the model's performance; a single block cannot effectively learn feature extraction, while too many blocks can risk model overfitting \cite{chen2022spatio}. Including LSTM with variational dropout or VDO helps regularize the sequential model; without it, the model becomes sensitive to overfitting \cite{gal2016theoretically}. BCNN involves parallel convolutional branches. One is a regular 1D-convolution, capturing local and global input data information. The other is a 1 x 1 convolution, providing channel-wise adaptation and dimensionality reduction, allowing the network to focus on the most informative channels. By having parallel branches, the model can simultaneously learn features at multiple scales, improving its ability to understand local and global contexts in the data.

We assume that the optimal ApSense configuration also depends on the characteristics of the datasets. Hence, we demonstrate this point by adjusting specific parameters: $M$ (with value of 1, 2, 3, 4, and 5), VDO (+/-), and BCNN (+/-) while exploring both MESA and HeartBEAT datasets. (+/-) implies the presence (+) or absence (-) of the function or block from the configuration.

\subsection{Experiment II: Benchmarking ApSense Against SOTA Algorithms in OSA Onset Recognition Task}

After experimenting with configuring ApSense across two datasets in Experiment I, we select the optimal ApSense configuration per each dataset with the highest AUROC to evaluate the performance against the SOTA algorithms: RRWaveNet, PPGNetSA, AIOSA (unimodal variant), DRIVEN and LeNet-5.





\subsection{Experiment III: Overnight AHI Estimation}\label{Overnight AHI Estimation}
Clinicians and researchers in sleep studies categorize the severity of OSA by employing predetermined AHI thresholds as the standard clinical guideline. The conventional AHI calculation is time-consuming, requiring the doctor to manually annotate events throughout the night. This experiment demonstrates the feasibility of applying ApSense with the optimal configuration reported in the earlier experiments to automatically recognize individual OSA events and accumulate them as total events. By dividing the total events by the number of sleep hours at night, we can estimate a predicted AHI or pAHI as quantified in equation \eqref{pAHI}.

\begin{equation}\label{pAHI}
    \text{pAHI} = \frac{\text{number of apneic predictions}}{30 \times \text{number of total windows} + 30} \times 3600 
\end{equation}
Since the ApSense's performance has been calculated in an event-based manner, in this experiment, we have to calculate the ground truth of overnight AHI or sAHI for one subject by accumulating total OSA events annotated according to the clinical guidelines and dividing them against the entire sleeping time of the interested night. As explained in the data preparation section, considering that two consecutive 60-second windows overlap by 50\% or 30 seconds, we define sAHI per hour as expressed in equation \eqref{sAHI},
\begin{equation}\label{sAHI}
    \text{sAHI} = \frac{\text{number of apneic windows}}{30 \times \text{number of total windows} + 30} \times 3600 
\end{equation}
where a prediction is made per window. Ideally, we would have $\text{pAHI} = \text{sAHI}$ when the model always makes a correct prediction. We present the benefits of pAHI compared to the ground truth sAHI by reporting correlation metrics from three methods: Pearson correlation, Spearman's rank correlation, and Kendall's Tau, ensuring reliability.

%% file: content/Results.tex
\section{Results}
\label{sec:results}

\subsection{ApSense Configuration}

As introduced in Section \ref{sec:med} and \autoref{fig:model2}, ApSense consists of three essential components: the spatiotemporal blocks constructed from the cascading RVarDSepBlocks with different filter sizes, the BCNN, and the FC layers. In this experiment, we optimize ApSense by configuring some hyperparameters of these components, including $M$, VDO, and BCNN. \autoref{table:exp_prove} display the results of this study.

The number of the stacked RVarDSepBlocks for each feature map size ($M$) shows greater significance on the HeartBEAT dataset. Applying fewer or more than four blocks leads to a noticeable decline in the ApSense's performance, whereas such impact is not shown on the MESA dataset. The results indicate that the optimal value of $M$ differs between the two datasets. However, configuring $M=4$ on both datasets allows the best AUROC to be achieved by toggling VDO or BCNN.

For the MESA dataset, ApSense achieves the highest AUROC without the BCNN. It outperforms the default ApSense with approximate increases of 3\%pt. for accuracy, 4\%pt. for F1 score, 8\%pt. for sensitivity, 2\%pt. for specificity, and 6\%pt. for AUROC.

On the contrary, for the HeartBEAT dataset, there is a tradeoff between specificity and sensitivity without the BCNN. We address this tradeoff by reintroducing the BCNN in addition to VDO, yielding an approximate 7\%pt. improvement in AUROC. Incorporating only the BCNN is not as effective against the tradeoff as having both components.

\begin{table*}
\centering
\tiny
\caption{Ablation study on ApSense: certain components are optimized or removed for understanding each component's contribution to the overall system. Metrics are reported in average and standard deviation (SD) percentages over five folds.}
\label{Ablation study}
\resizebox{\textwidth}{!}{%
    \input{new_tables/Ablationstudy}

    }

\begin{tablenotes}
  \scriptsize
  \item  \textbf{Abbreviations:} $M$: Number of stacked RVarDsepBlocks for different filter sizes, as illustrated in \autoref{fig:model2}. VDO: LSTM variational dropout. BCNN: Branch convolutional neural network.
\end{tablenotes}
\end{table*}


\begin{table*}
\centering
\tiny
\caption{Classification results for OSA event recognition, comparing the optimal configuration of ApSense to baseline models on the two datasets. The results are average percentage $\pm$ standard deviation (SD) over five folds.}
\label{Experiment_I}
\resizebox{\textwidth}{!}{%
    \input{new_tables/Experiment_I_revised}

    }
\end{table*}

\subsection{Benchmarking ApSense Against SOTA Algorithms in OSA Onset Recognition Task}
As reported in \autoref{Experiment_I}, we evaluate the algorithms' performance by validating in a subject-independent scheme that does not involve any data from the new users for model training. We select the best configuration of ApSense that yields the best AUROC score from \autoref{Ablation study} and compare it with the SOTA algorithms. ApSense is superior to other baseline methods on the MESA dataset, except for DRIVEN. In higher variance datasets such as HearBEAT, ApSense can balance the sensitivity and specificity tradeoff, while other algorithms are biased toward either sensitivity or specificity. This is consistent with ApSense, which also gives the highest AUROC in the HeartBEAT dataset.

\subsection{Overnight AHI Estimation}
We further demonstrate the feasibility of ApSense in a practical scheme such as AHI estimation. In this study, estimated AHI from ApSense is calculated from overnight OSA onset recognition and reported as predicted AHI (pAHI). This estimation divides the total predicted OSA onsets by the entire sleep duration. We compare pAHI to the ground truth, sAHI, to evaluate the estimation performance. Detailed derivations of pAHI and sAHI are explained in Section \ref{Overnight AHI Estimation}.

\autoref{Experiment_corre} demonstrates the reliability of AHI estimation through the correlation metrics between sAHI and pAHI from three different methods. Similarly, DRIVEN scores the best on the MESA dataset, while ApSense performs the best on the HeartBeat dataset.



\begin{table}
\centering
\caption{Comparison of correlation coefficients between predicted AHI (pAHI) and ground truth sampled AHI (sAHI) across different methods}
\label{Experiment_corre}
\resizebox{\columnwidth}{!}{%
    \input{new_tables/correlation_results}
    }
\end{table}

%% file: new_tables/Ablationstudy.tex
\begin{tabular}{@{}lcccccccc@{}}
\toprule[0.2em]
\multirow{2}{*}{\textbf{Dataset}} & \multirow{2}{*}{$\bm{M}$} 
& \multirow{2}{*}{\textbf{VDO}}
& \multirow{2}{*}{\textbf{BCNN}}
& \multicolumn{5}{c}{\textbf{Metrics}} \\
\cmidrule[0.1em]{5-9}
& & & & Accuracy & Macro F1 & Sensitivity & Specificity & AUROC\\ 

\midrule[0.1em]

\multirow{7}{*}{\emph{MESA}}

& 4     &\ding{51}   &\ding{53}     &\textbf{75.10 $\pm$ 2.70}	&\textbf{65.73 $\pm$ 4.84}	&\textbf{61.71 $\pm$ 17.41}	&\textbf{78.46 $\pm$ 1.62}	&\textbf{77.64 $\pm$ 7.93} \\

& 4     &\ding{53}   &\ding{51}     &70.96 $\pm$ 1.35	&61.82 $\pm$ 0.90	&57.86 $\pm$ 4.05	&74.08 $\pm$ 2.11	&72.28 $\pm$ 2.12 \\

& 4     &\ding{51}   &\ding{51}     &71.83 $\pm$ 0.33	&61.71 $\pm$ 0.60	&53.65 $\pm$ 3.86	&76.15 $\pm$ 1.21	&71.69 $\pm$ 1.95 \\

& 1     &\ding{51}   &\ding{51}     &69.73 $\pm$ 1.27	&61.54 $\pm$ 0.47	&61.96 $\pm$ 4.37	&71.56 $\pm$ 2.42	&72.82 $\pm$ 1.53 \\

& 2     &\ding{51}   &\ding{51}     &72.15 $\pm$ 1.38	&62.14 $\pm$ 1.09	&54.30 $\pm$ 2.15	&76.35 $\pm$ 1.66	&71.90 $\pm$ 1.78\\

& 3     &\ding{51}   &\ding{51}     &72.40 $\pm$ 1.50	&62.23 $\pm$ 0.65	&53.79 $\pm$ 2.80	&76.78 $\pm$ 2.06	&72.26 $\pm$ 1.28 \\

& 5     &\ding{51}   &\ding{51}     &70.18 $\pm$ 2.14	&60.95 $\pm$ 1.43	&56.53 $\pm$ 3.70	&73.48 $\pm$ 3.49	&71.22 $\pm$ 1.22 \\

\midrule[0.1em]

\multirow{7}{*}{\emph{HeartBEAT}} 

& 4     &\ding{51}   &\ding{53}     &76.94 $\pm$ 2.42	&54.20 $\pm$ 0.72	&24.20 $\pm$ 3.84	&85.09 $\pm$ 2.68	&67.07 $\pm$ 1.16 \\

& 4     &\ding{53}   &\ding{51}     &63.72 $\pm$ 1.96	&52.80 $\pm$ 1.16	&58.19 $\pm$ 2.47	&64.59 $\pm$ 2.37	&68.34 $\pm$ 2.16 \\

& 4     &\ding{51}   &\ding{51}     &\textbf{62.31 $\pm$ 3.06}	&\textbf{54.12 $\pm$ 3.55}	&\textbf{74.87 $\pm$ 10.91}	&\textbf{60.26 $\pm$ 3.35}	&\textbf{74.35 $\pm$ 6.46} \\

& 1     &\ding{51}   &\ding{51}   &58.71 $\pm$ 2.93	&50.65 $\pm$ 1.25	&68.50 $\pm$ 3.81	&57.10 $\pm$ 3.92	&68.04 $\pm$ 1.94 \\

& 2     &\ding{51}   &\ding{51}  &61.71 $\pm$ 2.19	&52.12 $\pm$ 1.06	&63.28 $\pm$ 2.81	&61.43 $\pm$ 2.68	&68.40 $\pm$ 1.98 \\

& 3     &\ding{51}   &\ding{51}     &63.54 $\pm$ 4.90	&52.33 $\pm$ 2.02	&56.87 $\pm$ 8.55	&64.50 $\pm$ 6.89	&67.66 $\pm$ 2.35  \\

& 5     &\ding{51}   &\ding{51}     &60.99 $\pm$ 11.04	&49.85 $\pm$ 2.62	&62.46 $\pm$ 26.23	&60.71 $\pm$ 17.01	&68.25 $\pm$ 2.09 \\

\bottomrule[0.2em]
\end{tabular}

\label{table:exp_prove}

%% file: new_tables/Experiment_I_revised.tex
{\renewcommand{\arraystretch}{1.00}
\begin{tabular}{@{}lllccccc@{}}

\toprule[0.2em]
\multirow{2}{*}{\textbf{Dataset}} & \multirow{2}{*}{\textbf{Model}} & 
\multirow{2}{*}{\textbf{\shortstack[l]{Trainable \\ Params}}} & 
\multicolumn{4}{c}{\textbf{Metrics}} \\
\cmidrule[0.1em]{4-8}
& & & Accuracy & Macro F1 & Sensitivity & Specificity &AUROC\\ 

\midrule[0.1em]

\multirow{6}{*}{\emph{MESA}}

& RRWaveNet & 6,969,008 &69.81 $\pm$ 0.84 &61.44 $\pm$ 0.45 &61.09 $\pm$ 4.88	&71.89 $\pm$ 2.11	&71.87 $\pm$ 1.53 \\

& PPGNetSA  & 66,753 & 71.39 $\pm$ 1.11	&61.94 $\pm$ 1.36	&56.56 $\pm$ 3.50	&74.93 $\pm$ 1.56	&72.13 $\pm$ 2.01 \\

&AIOSA (Unimodal variant)  & 168,965 &71.86 $\pm$ 1.43	&64.52 $\pm$ 0.88	&63.46 $\pm$ 7.35  &73.85 $\pm$ 3.24	&75.88 $\pm$ 2.36 \\

&LeNet-5  & 13,569 & 67.60 $\pm$ 6.81  &54.88 $\pm$ 5.09  &50.80 $\pm$ 25.61	&71.49 $\pm$ 14.34	&64.62 $\pm$ 7.44 \\

&DRIVEN & 20,715,078 & 73.93 $\pm$ 2.74 &\textbf{67.48 $\pm$ 3.08} &\textbf{77.17 $\pm$ 8.63} &73.26 $\pm$ 2.37	&\textbf{82.45 $\pm$ 5.40} \\

& ApSense   & 257,611  & \textbf{75.10 $\pm$ 2.70}	&65.73 $\pm$ 4.84	&61.71 $\pm$ 17.41 & \textbf{78.46 $\pm$ 1.62}	&77.64 $\pm$ 7.93 \\

\midrule[0.1em]

\multirow{6}{*}{\emph{HeartBEAT}} 

&RRWaveNet  & 6,969,008 &51.33 $\pm$ 3.45	& 47.04 $\pm$ 2.17  &85.76 $\pm$ 3.92	 &45.96 $\pm$ 4.49	 &69.46 $\pm$ 2.23 \\

&PPGNetSA  & 66,753 & 54.78 $\pm$ 2.64	 &49.09 $\pm$ 1.55	&79.77 $\pm$ 2.03	&50.86 $\pm$ 3.31	&69.28 $\pm$ 1.82 \\

&AIOSA (Unimodal variant)  & 168,965 &47.95 $\pm$ 2.34	&45.01 $\pm$ 1.65	&\textbf{92.95 $\pm$ 2.17}	&40.49 $\pm$ 2.91	& 70.20 $\pm$ 2.15 \\

&LeNet-5  & 13,569 & 60.74 $\pm$ 3.67	&51.70 $\pm$ 1.49	&65.55 $\pm$ 5.49 &59.90 $\pm$ 4.98	&68.03 $\pm$ 1.77 \\

&DRIVEN  & 20,715,078 & 55.52 $\pm$ 1.44	&49.13 $\pm$ 1.04	&74.85 $\pm$ 3.66	&52.53 $\pm$ 1.37	&67.54 $\pm$ 2.32 \\

& ApSense  & 201,411 &\textbf{62.31 $\pm$ 3.06}	 &\textbf{54.12 $\pm$ 3.55}	&74.87 $\pm$ 10.91	&\textbf{60.26 $\pm$ 3.35}	&\textbf{74.35 $\pm$ 6.46} \\

\bottomrule[0.2em]
\end{tabular}

}

%% file: new_tables/correlation_results.tex
\begin{tabular}{@{}ccccc@{}}
\toprule[0.2em]
\multicolumn{1}{l}{\textbf{Dataset}} &\textbf{Model} &\textbf{Pearson} & \textbf{Spearman’s rank} & \textbf{Kendall’s Tau}
\\ \midrule[0.1em]
    
    \multirow{6}{*}{MESA} 
    
    &RRWaveNet
    &0.393  
    &0.468 
    &0.330	\\
     
    &PPGNetSA 
    &0.308
    &0.329	
    &0.222\\
    
    &AIOSA (Unimodal variant)
    &0.399	
    &0.446
    &0.315 \\

    &LeNet-5 
    &0.160 
    &0.204	
    &0.143 \\

    &DRIVEN 
    &\textbf{0.596} 
    &\textbf{0.607}	
    &\textbf{0.429} \\

    &ApSense
    &0.532	
    &0.533
    &0.391 \\

    \midrule[0.1em]
     
    \multirow{6}{*}{HeartBeat} 
    
    &RRWaveNet
    &0.509  
    &0.583
    &0.417	\\
     
    &PPGNetSA 
    &0.515
    &0.569
    &0.410\\
    
    &AIOSA (Unimodal variant)
    &0.578	
    &0.630
    &0.450 \\

    &LeNet-5 
    & 0.461
    & 0.531	
    & 0.372 \\

    &DRIVEN 
    &0.397 
    &0.433	
    &0.305 \\

    &ApSense
    &\textbf{0.642}	
    &\textbf{0.650}
    &\textbf{0.469}  \\

    \bottomrule[0.2em]\\

\end{tabular}

%% file: content/Discussion.tex
\section{Discussion}
\label{sec:discussion}

\begin{figure}

  \center
  \includegraphics[width=\columnwidth]{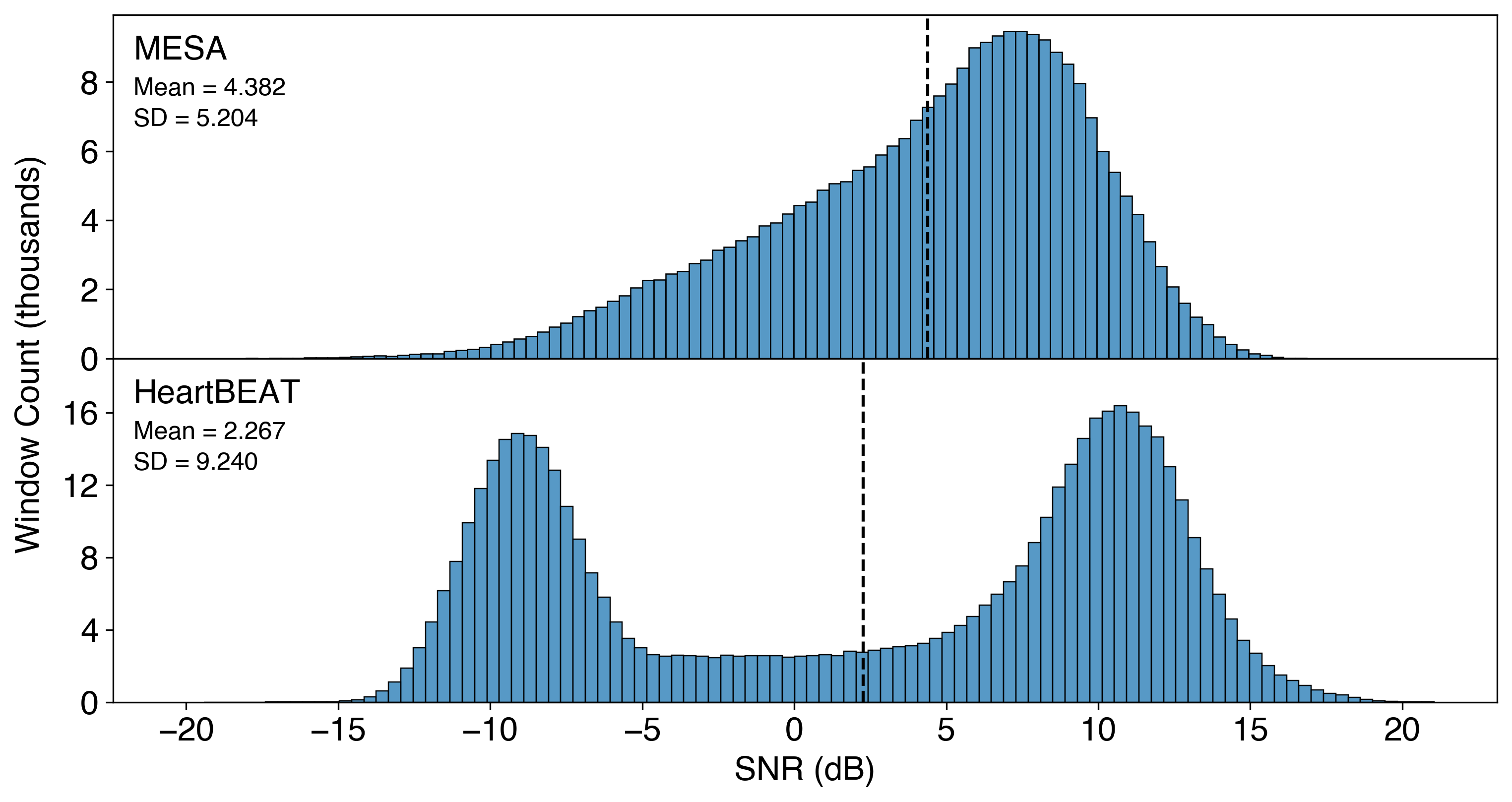}
  \caption{Signal-to-noise ratio of all signal windows before feature extraction, computed using the same method in \cite{10098530}. Unlike MESA, the SNR distribution of PPG windows from the HeartBEAT dataset shows bimodality with a smaller mode of negative SNR, implying many PPG windows with low signal quality.}
  \label{fig:snr}
\end{figure}

\subsection{Differences between the Two Benchmark Datasets}
Although DRIVEN yields the best performance on the MESA dataset, it requires a substantial amount of the trainable parameters (20.7 million), approximately 80 times greater than that of ApSense (257,611), as shown in \autoref{Experiment_I}. DRIVEN also struggles to learn effectively on the HeartBEAT dataset, where ApSense demonstrates the highest efficacy. As shown in \autoref{fig:snr}, the HeartBEAT dataset displays bimodality with a minor mode exhibiting negative SNR, indicating that many PPG windows have lower signal quality than the MESA dataset. This suggests that DRIVEN might overfit to noise and specific patterns in the training data, failing to generalize well to unseen data. In contrast, ApSense offers a more generalized approach to learning in high-variance datasets. The tradeoff between model size and performance is critical in resource-constrained and noisy data in real-world environments such as wearable devices. ApSense, with its smaller architecture, requires fewer resources, ensuring faster data processing, longer battery life, and robustness in operation.

\subsection{ApSense's Characteristics}
Confronting challenges in two dissimilar datasets, ApSense excels while other methods falter, notably in balancing sensitivity and specificity while optimizing AUROC. ApSense starts with spatiotemporal blocks, namely RVarDSepBlocks, which simultaneously capture complex spatial patterns and temporal dependencies. Integrating LSTM with a variational dropout layer inside RVarDSepBlocks also enhances sensitivity, which is crucial in diagnostic procedures prioritizing the identification of true positives. This dropout aids the model in handling uncertainty in LSTM connections, boosting reliability in noisy data scenarios. Moreover, an optimal number of RVarDSepBlocks is essential - too few may not be able to extract the high-level spatial-temporal information effectively, and too many may cause overfitting and not learning the necessary features well.

In the case of working with a high variance dataset, once the transformed data exits the RVarDSepBlocks, the proposed branch CNN block or the BCNN consequently captures local and global information and prevents over-reliance on specific patterns. 

In summary, the optimal configuration of ApSense depends on the dataset characteristics. To bolster ApSense's flexibility, ApSense could benefit from training using different datasets, enabling a generalization over a diverse range of data. We pose this as a high-impact research challenge that requires extensive work in the future.

\subsection{Overnight AHI Estimation}
According to the experimental results, ApSense shows the highest reliability in predicting AHI, even with high-variance data such as the HeartBEAT dataset, demonstrating the feasibility of using ApSense as a future tool for automatic OSA severity monitoring, overcoming the limitations of obtrusive and expensive clinical procedures which are unsuitable for long-term OSA monitoring. 

This application shows a high potential in practically enhancing automatic OSA monitoring, so further validation should be done. We must conduct a simultaneous assessment using a medical staff and our proposed method on patients for the AHI estimation. An intensive full-night test is recommended, especially for symptomatic individuals.



%% file: content/Conclusion.tex
\section{Conclusion}
\label{sec:con}

Compared to the existing algorithms, ApSense, using domain knowledge-based feature extraction and tailored deep neural architecture, gives promising results in OSA event recognition on two large-scale datasets widely known for sleep research. Our ablation study further emphasizes the importance and tailors a strategic optimization of each deep neural network component of ApSense to tackle each dataset. Moreover, our experimental setup proves that applying ApSense in remote health scenarios or home uses such as OSA pre-screening is feasible. Yet, full-night AHI estimation should be validated against the clinician's inspection for future use in the clinical aspect.

%% file: bare_jrnl_new_sample4.bbl
\begin{thebibliography}{10}
\providecommand{\url}[1]{#1}
\csname url@samestyle\endcsname
\providecommand{\newblock}{\relax}
\providecommand{\bibinfo}[2]{#2}
\providecommand{\BIBentrySTDinterwordspacing}{\spaceskip=0pt\relax}
\providecommand{\BIBentryALTinterwordstretchfactor}{4}
\providecommand{\BIBentryALTinterwordspacing}{\spaceskip=\fontdimen2\font plus
\BIBentryALTinterwordstretchfactor\fontdimen3\font minus
  \fontdimen4\font\relax}
\providecommand{\BIBforeignlanguage}[2]{{%
\expandafter\ifx\csname l@#1\endcsname\relax
\typeout{** WARNING: IEEEtran.bst: No hyphenation pattern has been}%
\typeout{** loaded for the language `#1'. Using the pattern for}%
\typeout{** the default language instead.}%
\else
\language=\csname l@#1\endcsname
\fi
#2}}
\providecommand{\BIBdecl}{\relax}
\BIBdecl

\bibitem{banno2007sleep}
K.~Banno and M.~H. Kryger, ``Sleep apnea: clinical investigations in humans,''
  \emph{Sleep medicine}, vol.~8, no.~4, pp. 400--426, 2007.

\bibitem{benjafield2019estimation}
A.~V. Benjafield, N.~T. Ayas, P.~R. Eastwood, R.~Heinzer, M.~S. Ip, M.~J.
  Morrell, C.~M. Nunez, S.~R. Patel, T.~Penzel, J.-L. P{\'e}pin \emph{et~al.},
  ``Estimation of the global prevalence and burden of obstructive sleep apnoea:
  a literature-based analysis,'' \emph{The Lancet Respiratory Medicine},
  vol.~7, no.~8, pp. 687--698, 2019.

\bibitem{9847226}
Q.~Shen, X.~Yang, L.~Zou, K.~Wei, C.~Wang, and G.~Liu, ``Multitask residual
  shrinkage convolutional neural network for sleep apnea detection based on
  wearable bracelet photoplethysmography,'' \emph{IEEE Internet of Things
  Journal}, vol.~9, no.~24, pp. 25\,207--25\,222, 2022.

\bibitem{ALLEN2022189}
J.~Allen, ``6 - photoplethysmography for the assessment of peripheral vascular
  disease,'' in \emph{Photoplethysmography}, J.~Allen and P.~Kyriacou,
  Eds.\hskip 1em plus 0.5em minus 0.4em\relax Academic Press, 2022, pp.
  189--235.

\bibitem{10.5665/sleep.4732}
X.~Chen, R.~Wang, P.~Zee, P.~L. Lutsey, S.~Javaheri, C.~Alcántara, C.~L.
  Jackson, M.~A. Williams, and S.~Redline, ``{Racial/Ethnic Differences in
  Sleep Disturbances: The Multi-Ethnic Study of Atherosclerosis (MESA)},''
  \emph{Sleep}, vol.~38, no.~6, pp. 877--888, 06 2015.

\bibitem{LEWIS201759}
E.~F. Lewis, R.~Wang, N.~Punjabi, D.~J. Gottlieb, S.~F. Quan, D.~L. Bhatt,
  S.~R. Patel, R.~Mehra, R.~S. Blumenthal, J.~Weng, M.~Rueschman, and
  S.~Redline, ``Impact of continuous positive airway pressure and oxygen on
  health status in patients with coronary heart disease, cardiovascular risk
  factors, and obstructive sleep apnea: A heart biomarker evaluation in apnea
  treatment (heartbeat) analysis,'' \emph{American Heart Journal}, vol. 189,
  pp. 59--67, 2017.

\bibitem{bernardini2022osasud}
A.~Bernardini, A.~Brunello, G.~L. Gigli, A.~Montanari, and N.~Saccomanno,
  ``Osasud: A dataset of stroke unit recordings for the detection of
  obstructive sleep apnea syndrome,'' \emph{Scientific Data}, vol.~9, no.~1, p.
  177, 2022.

\bibitem{gonzalo2019}
G.~C. Gutiérrez-Tobal, D.~Álvarez, A.~Crespo, F.~del Campo, and R.~Hornero,
  ``Evaluation of machine-learning approaches to estimate sleep apnea severity
  from at-home oximetry recordings,'' \emph{IEEE Journal of Biomedical and
  Health Informatics}, vol.~23, no.~2, pp. 882--892, 2019.

\bibitem{remo2020}
R.~Lazazzera, M.~Deviaene, C.~Varon, B.~Buyse, D.~Testelmans, P.~Laguna,
  E.~Gil, and G.~Carrault, ``Detection and classification of sleep apnea and
  hypopnea using ppg and spo$_2$ signals,'' \emph{IEEE Transactions on
  Biomedical Engineering}, vol.~68, no.~5, pp. 1496--1506, 2021.

\bibitem{chen2021}
Y.~Chen, W.~Wang, Y.~Guo, H.~Zhang, Y.~Chen, and L.~Xie, ``A single-center
  validation of the accuracy of a photoplethysmography-based smartwatch for
  screening obstructive sleep apnea,'' \emph{Nature and Science of Sleep},
  vol.~13, p. 1533, 2021.

\bibitem{bernardini2021aiosa}
A.~Bernardini, A.~Brunello, G.~L. Gigli, A.~Montanari, and N.~Saccomanno,
  ``Aiosa: An approach to the automatic identification of obstructive sleep
  apnea events based on deep learning,'' \emph{Artificial Intelligence in
  Medicine}, vol. 118, p. 102133, 2021.

\bibitem{li2023deep}
F.~Li, Y.~Xu, J.~Chen, P.~Lu, B.~Zhang, and F.~Cong, ``A deep learning model
  developed for sleep apnea detection: A multi-center study,'' \emph{Biomedical
  Signal Processing and Control}, vol.~85, p. 104689, 2023.

\bibitem{10098530}
P.~Osathitporn, G.~Sawadwuthikul, P.~Thuwajit, K.~Ueafuea, T.~Mateepithaktham,
  N.~Kunaseth, T.~Choksatchawathi, P.~Punyabukkana, E.~Mignot, and
  T.~Wilaiprasitporn, ``Rrwavenet: A compact end-to-end multiscale residual cnn
  for robust ppg respiratory rate estimation,'' \emph{IEEE Internet of Things
  Journal}, vol.~10, no.~18, pp. 15\,943--15\,952, 2023.

\bibitem{10.1093/sleep/22.5.667}
``{Sleep-Related Breathing Disorders in Adults: Recommendations for Syndrome
  Definition and Measurement Techniques in Clinical Research},'' \emph{Sleep},
  vol.~22, no.~5, pp. 667--689, 08 1999.

\bibitem{10.1093/jamia/ocy064}
G.-Q. Zhang, L.~Cui, R.~Mueller, S.~Tao, M.~Kim, M.~Rueschman, S.~Mariani,
  D.~Mobley, and S.~Redline, ``{The National Sleep Research Resource: towards a
  sleep data commons},'' \emph{Journal of the American Medical Informatics
  Association}, vol.~25, no.~10, pp. 1351--1358, 05 2018.

\bibitem{dumitrache2013role}
S.~Dumitrache-Rujinski, G.~Calcaianu, D.~Zaharia, C.~L. Toma, and M.~Bogdan,
  ``The role of overnight pulse-oximetry in recognition of obstructive sleep
  apnea syndrome in morbidly obese and non obese patients,'' \emph{Maedica},
  vol.~8, no.~3, p. 237, 2013.

\bibitem{2020SciPy-NMeth}
P.~Virtanen, R.~Gommers, T.~E. Oliphant, M.~Haberland, T.~Reddy, D.~Cournapeau,
  E.~Burovski, P.~Peterson, W.~Weckesser, J.~Bright, S.~J. {van der Walt},
  M.~Brett, J.~Wilson, K.~J. Millman, N.~Mayorov, A.~R.~J. Nelson, E.~Jones,
  R.~Kern, E.~Larson, C.~J. Carey, {\.I}.~Polat, Y.~Feng, E.~W. Moore,
  J.~{VanderPlas}, D.~Laxalde, J.~Perktold, R.~Cimrman, I.~Henriksen, E.~A.
  Quintero, C.~R. Harris, A.~M. Archibald, A.~H. Ribeiro, F.~Pedregosa, P.~{van
  Mulbregt}, and {SciPy 1.0 Contributors}, ``{{SciPy} 1.0: Fundamental
  Algorithms for Scientific Computing in Python},'' \emph{Nature Methods},
  vol.~17, pp. 261--272, 2020.

\bibitem{delessert2010pulse}
A.~Delessert, F.~Espa, A.~Rossetti, G.~Lavigne, M.~Tafti, and R.~Heinzer,
  ``Pulse wave amplitude drops during sleep are reliable surrogate markers of
  changes in cortical activity,'' \emph{Sleep}, vol.~33, no.~12, pp.
  1687--1692, Dec. 2010.

\bibitem{GROTE2003141}
L.~Grote, D.~Zou, H.~Kraiczi, and J.~Hedner, ``Finger plethysmography—a
  method for monitoring finger blood flow during sleep disordered breathing,''
  \emph{Respiratory Physiology \& Neurobiology}, vol. 136, no.~2, pp. 141--152,
  2003, sleep and Breathing: from molecules to patient populations.

\bibitem{CYGANKIEWICZ2013379}
I.~Cygankiewicz and W.~Zareba, ``Chapter 31 - heart rate variability,'' in
  \emph{Autonomic Nervous System}, ser. Handbook of Clinical Neurology, R.~M.
  Buijs and D.~F. Swaab, Eds.\hskip 1em plus 0.5em minus 0.4em\relax Elsevier,
  2013, vol. 117, pp. 379--393.

\bibitem{Nakayama_2019}
C.~Nakayama, K.~Fujiwara, Y.~Sumi, M.~Matsuo, M.~Kano, and H.~Kadotani,
  ``Obstructive sleep apnea screening by heart rate variability-based
  apnea/normal respiration discriminant model,'' \emph{Physiological
  Measurement}, vol.~40, no.~12, p. 125001, dec 2019.

\bibitem{KHANDOKER2011204}
A.~H. Khandoker, C.~K. Karmakar, and M.~Palaniswami, ``Comparison of pulse rate
  variability with heart rate variability during obstructive sleep apnea,''
  \emph{Medical Engineering \& Physics}, vol.~33, no.~2, pp. 204--209, 2011.

\bibitem{10.1016/S0895-7061(01)02143-4}
J.~S. Loredo, S.~Ancoli-Israel, and J.~E. Dimsdale, ``{Sleep quality and blood
  pressure dipping in obstructive sleep apnea*},'' \emph{American Journal of
  Hypertension}, vol.~14, no.~9, pp. 887--892, 09 2001.

\bibitem{Suzuki1996-uf}
M.~Suzuki, C.~Guilleminault, K.~Otsuka, and T.~Shiomi,
  ``\BIBforeignlanguage{en}{Blood pressure ``dipping'' and ``non-dipping'' in
  obstructive sleep apnea syndrome patients},''
  \emph{\BIBforeignlanguage{en}{Sleep}}, vol.~19, no.~5, pp. 382--387, Jun.
  1996.

\bibitem{chen2022spatio}
J.~Chen, M.~Shen, W.~Ma, and W.~Zheng, ``A spatio-temporal learning-based model
  for sleep apnea detection using single-lead ecg signals,'' \emph{Frontiers in
  Neuroscience}, vol.~16, p. 972581, 2022.

\bibitem{szegedy2017inception}
C.~Szegedy, S.~Ioffe, V.~Vanhoucke, and A.~Alemi, ``Inception-v4,
  inception-resnet and the impact of residual connections on learning,'' in
  \emph{Proceedings of the AAAI conference on artificial intelligence},
  vol.~31, no.~1, 2017.

\bibitem{shi2023exploiting}
J.~Shi, Y.~Wang, Z.~Yu, G.~Li, X.~Hong, F.~Wang, and Y.~Gong, ``Exploiting
  multi-scale parallel self-attention and local variation via dual-branch
  transformer-cnn structure for face super-resolution,'' \emph{IEEE
  Transactions on Multimedia}, 2023.

\bibitem{retamales2024towards}
G.~Retamales, M.~E. Gavidia, B.~Bausch, A.~N. Montanari, A.~Husch, and
  J.~Goncalves, ``Towards automatic home-based sleep apnea estimation using
  deep learning,'' \emph{npj Digital Medicine}, vol.~7, no.~1, p. 144, 2024.

\bibitem{wang2019sleep}
T.~Wang, C.~Lu, G.~Shen, and F.~Hong, ``Sleep apnea detection from a
  single-lead ecg signal with automatic feature-extraction through a modified
  lenet-5 convolutional neural network,'' \emph{PeerJ}, vol.~7, p. e7731, 2019.

\bibitem{choirunnisa2018hybrid}
S.~Choirunnisa and J.~Lianto, ``Hybrid method of undersampling and oversampling
  for handling imbalanced data,'' in \emph{2018 International Seminar on
  Research of Information Technology and Intelligent Systems (ISRITI)}.\hskip
  1em plus 0.5em minus 0.4em\relax IEEE, 2018, pp. 276--280.

\bibitem{gal2016theoretically}
Y.~Gal and Z.~Ghahramani, ``A theoretically grounded application of dropout in
  recurrent neural networks,'' \emph{Advances in neural information processing
  systems}, vol.~29, 2016.

\end{thebibliography}
